\documentclass[sigconf]{acmart}
\usepackage{subfigure} 
\usepackage[ruled]{algorithm2e}
\usepackage{algorithmic}
\usepackage{threeparttable}
\usepackage{wasysym}
\usepackage{enumitem}
\usepackage{multirow}
\AtBeginDocument{%
  }

\setcopyright{acmlicensed}
\copyrightyear{2025}
\acmYear{2025}
\acmConference[SIGIR '25] {Proceedings of the 48th International ACM SIGIR Conference on Research and Development in Information Retrieval}{July 13--18, 2025}{Padua, Italy}
\acmBooktitle{Proceedings of the 48th International ACM SIGIR Conference on Research and Development in Information Retrieval (SIGIR '25), July 13--18, 2025, Padua, Italy}
\acmISBN{979-8-4007-1592-1/25/07}
\acmDOI{10.1145/3726302.3730098}

\begin{document}

\title{Towards Distribution Matching between Collaborative and Language Spaces for Generative Recommendation}

\author{Yi Zhang}
\authornote{The work was done while the author was visiting the University of Queensland.}
\affiliation{%
  \institution{Anhui University}
  \city{Hefei}
  \country{China}}
\email{zhangyi.ahu@gmail.com}

\author{Yiwen Zhang}
\authornote{Yiwen Zhang and Hongzhi Yin are co-corresponding authors.}
\affiliation{%
  \institution{Anhui University}
  \city{Hefei}
  \country{China}
}
\email{zhangyiwen@ahu.edu.cn}

\author{Yu Wang}
\affiliation{%
  \institution{Anhui University}
  \city{Hefei}
  \country{China}
}
\email{wangyuahu@stu.ahu.edu.cn}

\author{Tong Chen}
\affiliation{%
  \institution{The University of Queensland}
  \city{Brisbane}
  \country{Australia}
}
\email{tong.chen@uq.edu.au}

\author{Hongzhi Yin}
\authornotemark[2]
\affiliation{%
  \institution{The University of Queensland}
  \city{Brisbane}
  \country{Australia}
}
\email{h.yin1@uq.edu.au}

\renewcommand{\shortauthors}{Yi Zhang, Yiwen Zhang, Yu Wang, Tong Chen, \& Hongzhi Yin}

\begin{abstract}
Generative recommendation aims to learn the underlying generative process over the entire item set to produce recommendations for users. Although it leverages non-linear probabilistic models to surpass the limited modeling capacity of linear factor models, it is often constrained by a trade-off between representation ability and tractability. With the rise of a new generation of generative methods based on pre-trained language models (LMs), incorporating LMs into general recommendation with implicit feedback has gained considerable attention. However, adapting them to generative recommendation remains challenging. The core reason lies in the mismatch between the input-output formats and semantics of generative models and LMs, making it challenging to achieve optimal alignment in the feature space. This work addresses this issue by proposing a model-agnostic generative recommendation framework called \textsf{DMRec}, which introduces a probabilistic meta-network to bridge the outputs of LMs with user interactions, thereby enabling an equivalent probabilistic modeling process. Subsequently, we design three cross-space distribution matching processes aimed at maximizing shared information while preserving the unique semantics of each space and filtering out irrelevant information. 
We apply \textsf{DMRec} to three different types of generative recommendation methods and conduct extensive experiments on three public datasets. The experimental results demonstrate that \textsf{DMRec} can effectively enhance the recommendation performance of these generative models, and it shows significant advantages over mainstream LM-enhanced recommendation methods.
\end{abstract}





\maketitle

\section{Introduction}
General recommender system \cite{ricci2011introduction} explicitly models user behavior patterns and preference rules by learning interactions between users and items \cite{rendle2009bpr}. The goal is to identify the optimal metric between users and items in the feature space, thereby fulfilling the requirements of collaborative filtering \cite{sarwar2001item, he2017neural} and establishing connections based solely on user or item IDs \cite{yuan2023go}. It is evident that interactions are crucial for establishing collaborative signals \cite{he2020lightgcn}, especially in matrix factorization \cite{rendle2009bpr} or neural network-based discriminative methods \cite{he2017neural, he2020lightgcn} that rely on modeling unique user and item embeddings \cite{chen2022thinking}. However, it is often challenging for platforms to obtain sufficient interactions, which raises a new concern: when interactions are extremely sparse, the effectiveness of recommender systems may be significantly compromised. 

Therefore, another avenue of exploration is generative recommendation \cite{yu2019vaegan, wang2017irgan, liang2018variational, wang2023diffusion}, which seeks to establish a comprehensive preference distribution for users, enabling the generation of unknown interactions based on known data points. For example, the widely used Variational Autoencoder (VAE) \cite{kingma2013auto, liang2018variational} attempts to construct an approximate variational distribution from limited interactions. This distribution typically consists of continuous random variables in a high-dimensional feature space, implicitly capturing the underlying patterns of user preferences. Going further, hierarchical VAE \cite{sonderby2016ladder} extends the distribution space into multiple hierarchies with the Markov process \cite{luo2022understanding}, making the input at each time step dependent on the output from the previous step, and eventually evolving into the well-known diffusion model \cite{ho2020denoising, wang2023diffusion}. The achievements of generative recommender systems are undoubtedly remarkable. However, as a fundamental paradigm of self-supervised learning \cite{liu2021self, yu2023self}, generative models are inherently limited to generating (or reconstructing) new samples that resemble the input \cite{kingma2013auto}, which is insufficient for recommendation tasks focused on predicting unknown interactions. Moreover, generative models are also constrained by the trade-off between model representation ability and tractability \cite{kingma2016improved}. Specifically, simplistic encoder-decoder structures may fail to capture the complexity of user preferences and may suffer from collapse phenomena \cite{wang2022posterior}, while more complex designs often make the approximate posterior difficult to handle, leading to unstable model training \cite{sohl2015deep, takahashi2019variational}.

With the rise of pre-trained Language Model (LM) \cite{touvron2023llama, zhao2023survey}, a new generation of generative models for text processing has been proposed in recent years and have achieved remarkable success across various domains \cite{chang2024survey}. In the recommendation domain, many groundbreaking works have similarly attempted to integrate auxiliary language models into recommendation to enhance the expressive power of the recommendation models. The first strategy is to integrate the recommendation process into the training of language models by fine-tuning the model's parameters \cite{liao2024llara, zhao2023survey} to adapt to the recommendation task \cite{geng2022recommendation, bao2023tallrec}. This strategy is often constrained by high training time costs \cite{ren2024representation} and may face limitations in certain scenarios \cite{liao2024llara} (\textit{e.g.}, next-item prediction \cite{kang2018self}). Another option is to treat the language model as an assistant for recommendation \cite{yuan2024fellas}. For example, works like KAR \cite{xi2024towards} and RLMRec \cite{ren2024representation} attempt to use language models to construct user (or item) profiles, relying on text embedding models \cite{neelakantan2022text} to map them into high-dimension feature spaces. Extensive practice \cite{xi2024towards, wei2024llmrec, ren2024representation} has proven the effectiveness of incorporating language models to assist recommendation models, but it is necessary to additionally consider the alignment of representations from different semantic spaces \cite{radford2021learning, ren2024representation}.

When we shift perspective to generative recommendation, the situation becomes significantly different. On the one hand, the generative recommendation process aims to produce a probability distribution over all items from interactions \cite{liang2018variational}, which is a point estimation process rather than modeling a unique embedding for each user or item \cite{he2020lightgcn}. In this case, there is a direct gap in that the text generated by language models does not directly reflect user's true distribution, and it is even difficult to express such text directly in a probabilistic form. Besides that, language models are not always as reliable as expected due to semantic noise \cite{wei2024llmrec} and hallucination \cite{ji2023survey}. The necessity of alignment has been widely pointed out in previous works \cite{radford2021learning, ren2024representation}. However, these embedding-based discriminative methods are often challenging to apply directly to probability distributions. When dealing with probability distributions, measuring their relationships is not merely a matter of calculating distances. It also requires considering more complex factors, such as probability density, overall distribution shapes, and inherent statistical properties \cite{xu2020learning, liu2022exploiting}. Moreover, our objective is not to align embeddings but to identify the optimal matching between probability distributions from different semantic spaces, thereby enabling lossless information transfer. 

% Therefore, when considering how to introduce a language model as an assistant to generative recommendation models, it will face the following two challenges:
% \begin{itemize}[leftmargin=*]
% \item[$\bullet$] How to model the language model in a coherent probabilistic format to assist in the training process of the generative model? 
% \item[$\bullet$] How to optimally match the probability distributions of the generative model and the language model for recommendation?
% \end{itemize}

To tackle the above challenges, we propose a novel Distribution Matching-based Framework for Generative Recommendation (\textbf{\textsf{DMRec}}), which can serve as a plug-and-play component for generative recommendation models. Specifically, we assume that the user's historical interactions and textual information originate from a collaborative space and a language space, respectively, and we model the user's preference distributions in each of these spaces: For the generative model in the collaborative space, we resort to variational inference \cite{kingma2013auto} to perform maximum likelihood estimation, thereby constructing an approximate variational distribution of users. For the language model in the language space, we first generate user (or item) profiles in a fixed format via system prompts and then convert them into fixed-length semantic vectors. To align the distributions in the collaborative space, we introduce a probabilistic meta-network that serves as a bridge between the two spaces, enabling consistent probabilistic modeling and dimensional transformations across both spaces. 
Subsequently, we propose three cross-space distribution matching strategies, aimed at maximizing shared information while preserving the unique semantics of each space and filtering out irrelevant information. All of the above processes are integrated into the \textsf{DMRec} framework. Since we do not restrict the encoder/decoder of the generative model in the collaborative space, nor the language model in the language space, \textsf{DMRec} can be considered a plug-and-play module, applicable to various generative recommendation models based on variational inference. The major contributions of this paper are summarized as follows:

\begin{itemize}[leftmargin=*]
\item[$\bullet$] We propose a model-agnostic generative recommendation framework called \textsf{DMRec}, which models the user's preference distributions in both the collaborative space and the language space.
\item[$\bullet$] We propose three cross-space distribution matching strategies to achieve a trade-off between maximizing shared information and preserving unique semantics from a distributional perspective.
\item[$\bullet$] We integrate into three types of generative recommendation models and conduct extensive experiments on three public datasets. The results demonstrate that \textsf{DMRec} not only significantly enhances the performance of the base models but also offers a clear advantage over other LM-based recommendation methods.
\end{itemize}

\section{Methodology}
\subsection{Problem Formulation}
\begin{figure*}[t]
\setlength{\abovecaptionskip}{0.1cm}
\setlength{\belowcaptionskip}{0.1cm} 
\centering
\includegraphics[width=\linewidth]{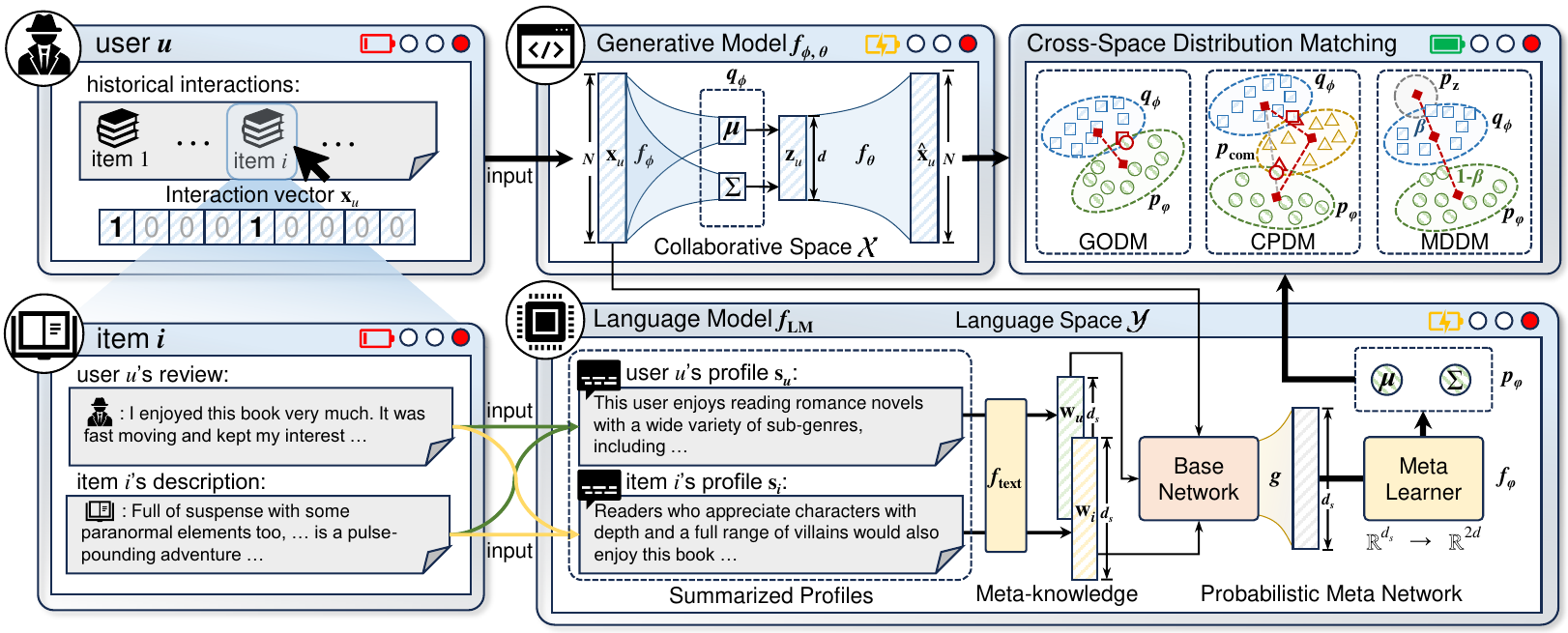}
\caption{
The proposed \textsf{DMRec} information flow, which models user preference distributions $q_\phi$ and $p_\varphi$ in the collaborative space $\mathcal X$ and language space $\mathcal Y$, respectively, and performs cross-space distribution matching through GODM, CPDM, or MDDM.}
\label{fig_model}
\end{figure*}

Without loss of generality, a general recommendation scenario contains $M$ users ($\mathcal U=\{u_1, u_2, ..., u_M\}$) and $N$ items ($\mathcal I=\{i_1, i_2, ..., i_N\}$). Based on existing works \cite{liang2018variational, he2020lightgcn}, given any user $u \in \mathcal U$, the historical interactions are stored as an interaction vector $\mathbf x_u \in \mathbb R^{1\times N}$, where if there is an observed interaction between user $u$ and item $i$, we then have $x_{ui}=1$. The task of the recommender system aims to learn the prediction model to predict user $u$'s preference score $\hat{x}_{uj}$ for all items $\{j \in \mathcal I /i\}$ that have not been interacted with. Based on this, we propose the model-agnostic generative recommendation framework \textsf{DMRec}, as illustrated in Fig. \ref{fig_model}.

\subsection{Distribution Modeling in Collaborative Space}
Given any user $u$, the historical interactions are represented as data point $\mathbf x_u$. For the generative model, the observed data point are modeled by the joint distribution $p(\mathbf x_u, \mathbf z_u)$, where $\mathbf z_u$ is an existing but unknown $d$-dimension continuous  variable for user $u$ in the collaborative space $\mathcal X \in \mathbb R^{d}$. Generative modeling aims to maximize the likelihood $p(\mathbf x_u)$ by directly marginalizing the latent variable $p(\mathbf x_u) =\int p(\mathbf x_u, \mathbf z_u)d\mathbf z_u $ \cite{luo2022understanding}. 
Intuitively, directly solving it is challenging, especially since we have no knowledge of the true nature for $\mathbf z_u$. An alternative attempt is to approximate the true posterior distribution via the Evidence Lower Bound (ELBO) \cite{kingma2013auto, luo2022understanding} as a  proxy objective to quantify the log-likelihood:
\begin{equation}
\label{likelihood}
\text{log}p(\mathbf x_u)=\text{log}\int p(\mathbf x_u, \mathbf z_u)d\mathbf z_u \ge \mathbb E_{q_{\phi}(\mathbf z_u|\mathbf x_u)}\left [\text{log}\frac{p(\mathbf x_u, \mathbf z_u)}{q_{\phi}(\mathbf z_u|\mathbf x_u)}    \right ],
\end{equation}
where $q_{\phi}(\mathbf z_u|\mathbf x_u)$ is an approximate posterior parameterized by $\phi$, which is conjugate to the prior belief $p(\mathbf z_u)$. Variational inference \cite{kingma2013auto, graves2011practical} is introduced in pioneering works \cite{li2017collaborative, liang2018variational}, aiming to optimize the optimal $q_{\phi}(\mathbf z_u|\mathbf x_u)$ amongst a family of posterior distributions parameterized by $\phi$. For the ELBO term, we have the following derivation:
\begin{equation}
\begin{aligned}
\label{elbo}
\text{log}p(\mathbf x_u) &\ge \mathbb E_{q_{\phi}(\mathbf z_u|\mathbf x_u)}\left [\text{log}\frac{p(\mathbf x_u, \mathbf z_u)}{q_{\phi}(\mathbf z_u|\mathbf x_u)}    \right ]\\
&= \mathbb E_{q_{\phi}(\mathbf z_u|\mathbf x_u)}[\text{log}p_{\theta}(\mathbf x_u|\mathbf z_u)] - \text{D}_{\text{KL}}(q_{\phi}(\mathbf z_u|\mathbf x_u)||p(\mathbf z_u)).
\end{aligned}
\end{equation}
The first term is the reconstruction term, which aims to ensure that the approximate distribution (viewed as the encoder parameterized by variational parameter $\phi$) can sample the correct latent vector $\mathbf z_u$, allowing the original data point $\mathbf x_u$ to be reconstructed (viewed as the decoder parameterized by generative parameter $\theta$). The second term is the reverse Kullback-Leibler (KL) divergence \cite{luo2022understanding}, which is considered as a regularization term for the variational parameter $\phi$ \cite{higgins2017beta, xu2020learning}, encouraging the approximate posterior $q_{\phi}(\mathbf z_u|\mathbf x_u)$ to be close to the standard prior belief $p(\mathbf z_u)$ \cite{liang2018variational}.

One of the most widespread applications of ELBO is the variational auto-encoder \cite{kingma2013auto}. As indicated by the central limit theorem, VAE assumes that the latent variables follow a multivariate Gaussian distribution with diagonal covariance:
\begin{equation}
\label{vae_q}
q_{\phi}(\mathbf z_u|\mathbf x_u)=\mathcal N\left (\mathbf z_u|\boldsymbol\mu_\phi(\mathbf x_u), \boldsymbol\Sigma_{\phi}(\mathbf x_u)\right),
\end{equation}
where $\boldsymbol\mu_\phi$ and $\boldsymbol\Sigma_{\phi}$ are the non-linear mean and covariance functions parameterized by $\phi$, respectively. When the covariance matrix is diagonal, we have $\boldsymbol\Sigma_{\phi}(\mathbf x_u)=\text{diag}[\boldsymbol\sigma^2_{\phi}(\mathbf x_u)]$, where $\boldsymbol\sigma_\phi$ is the standard deviation parameterized by $\phi$ \cite{xu2020learning}. 

\subsection{Distribution Modeling in Language Space}

The primary challenge encountered by recommender systems is the data sparsity \cite{yu2022graph, zhang2023revisiting}, which is typically mitigated through the incorporation of user and item attributes. By exploiting advanced text processing techniques and leveraging extensive domain-specific knowledge, a language model can significantly enrich the information of users and items, thereby constructing a distinctive and holistic feature profile \cite{ren2024representation}. Furthermore, it can distill complex information by applying inductive reasoning and generating concise summaries. Without loss of generality, given any language model, we can construct prompt templates governed by specific rules that define the format of input user and item attributes (\textit{i.e.}, title, tags, reviews, and descriptions), while ensuring that the language model generates concise and summarized profiles that adhere to predefined length constraints \cite{radford2021learning}:
\begin{equation}
\label{prompt}
\mathbf s_u=f_{\text{LM}}(\mathcal P(u)); \quad \mathbf s_i=f_{\text{LM}}(\mathcal P(i)),
\end{equation}
where $f_{\text{LM}}(\mathcal P(x))$ is the function that invokes the language model, taking as input a prompt $\mathcal P(x)$ that includes the attributes of $x$. The design of the prompt template is flexible and not the main focus of our paper. Therefore, we follow the design used in previous works \cite{xi2024towards, ren2024representation}, where the prompt $\mathcal P(i)$ for an item $i$ includes title, dataset-specific tags, and descriptions. For user $u$, the prompt $\mathcal P(u)$ contains descriptions and reviews of a sampled subset of items with which the user has interacted. Based on this, a text embedding model $f_\text{text}$ \cite{su2023one} is employed to transform the profile into a fixed-dimension semantic vector, which forms the basis for subsequent processing:
\begin{equation}
\label{text_embedding}
\mathbf w_u = f_\text{text}(\mathbf s_u); \quad \mathbf w_i=f_\text{text}(\mathbf s_i),
\end{equation}
where $\mathbf w_u \in \mathbb R^{1\times d_s}$ and $\mathbf w_i \in \mathbb R^{1\times d_s}$ are the 
$d_s$-dimension semantic vectors for user $u$ and item $i$ in the language space $\mathcal Y \in \mathbb R^{d_s}$, respectively. Both contain rich semantic information derived from the user and item profiles, and are thus considered as the meta-knowledge of the user $u$ and item $i$, respectively. 

The current challenges are mainly reflected in two aspects. Firstly, such semantic vectors fail to effectively capture the distribution of user preferences, especially as they cannot be represented as probability distributions. Secondly, these semantic vectors are not directly applicable to recommendation tasks, as the vectors generated by $f_\text{text}$ reside in a non-smooth anisotropic semantic space \cite{zhang2022effect} and may contain noise irrelevant to recommendation \cite{ren2024representation, sheng2024language}. It is important to note that our task is not to construct isotropic semantic representations through parameter whitening \cite{ermolov2021whitening}, but to establish collaborative signals \cite{he2020lightgcn} between language vectors and user interactions from a distributional perspective. Therefore, we propose a probabilistic meta network to enable semantic knowledge transformation, while fully considering the collaborative relations between user $u$ and interacted item set $\mathcal M(u)$:
\begin{equation}
\label{language_p}
\mathcal N(\boldsymbol{\mu}_{\varphi}, \boldsymbol{\Sigma}_{\varphi}) = f_\varphi \left (g(\mathbf x_u, \mathbf w_u, \{\mathbf w_i\}, i \in \mathcal M(u))\right ),
\end{equation}
where $f_\varphi: \mathbb{R}^{d_s} \to \mathbb{R}^{2d}$ is a meta-learner parameterized by 
$\varphi$, consisting of two fully connected layers with a Tanh activation function, while $g$ is the base network responsible for establishing the connection between the semantic vectors $\mathbf w_u$ and historical interactions $\mathbf x_u$. It has the following design:
\begin{equation}
\label{meta_network}
g(\mathbf x_u, \mathbf w_u, \{\mathbf w_i\}, i \in \mathcal M(u))=\underbrace{\mathbf W_{\mathcal I}^\top \mathbf x_u}_{\text{
Interacted items} }  + \underbrace{\mathbf w_u}_{\text{user bias} },
\end{equation}
where $\mathbf W_{\mathcal I} \in \mathbb R^{N \times d_s}$ is the meta-knowledge matrix of all items, and $\mathbf w_{u} \in \mathbb R^{1 \times d_s}$ is the meta-knowledge vector for user $u$, both of which are derived from Eq. \ref{text_embedding}. To match the collaborative space $\mathcal X$, we treat the output of $f_{\varphi}$ as the approximate posterior distribution of the user $u$ in the language space $\mathcal Y$, which includes the mean $\boldsymbol{\mu}_{\varphi}\in \mathbb R^{d}$ and covariance matrix $\boldsymbol{\Sigma}_{\varphi}\in \mathbb R^{d}$, \textit{i.e.}, $p_{\varphi}(\mathbf z_u|\mathbf s_u) \sim \mathcal N(\boldsymbol{\mu}_{\varphi}, \boldsymbol{\Sigma}_{\varphi})$, with sampling via the reparameterization trick \cite{kingma2013auto, luo2022understanding}.

\subsection{Cross-Space Distribution Matching}

For user $u$,  we have modeled the variational distribution $\mathcal N(\boldsymbol{\mu}_{\phi}, \boldsymbol{\Sigma}_{\phi})$ in the collaborative space $\mathcal X$ and the approximate posterior $\mathcal N(\boldsymbol{\mu}_{\varphi}, \boldsymbol{\Sigma}_{\varphi})$ in the language space $\mathcal Y$, respectively, both of which share the same distribution type and dimension. The most intuitive and easy-to-implement strategy is to directly exploit the distribution $\mathcal N(\boldsymbol{\mu}_{\varphi}, \boldsymbol{\Sigma}_{\varphi})$ in the language space $\mathcal Y$ for recommendation, or treat it as the prior in the collaborative space $\mathcal X$, optimizing the process in the collaborative space $\mathcal X$ by leveraging the additivity of two independent Gaussian distributions \cite{luo2022understanding, wang2023diffusion}.

Nonetheless, several challenges arise due to the discrepancy between the two spaces, including significant deviations in terms of probability density, distribution shape, and support regions \cite{xu2020learning}. Direct alignment may lead to irreversible  information distortion. Moreover, the distribution $\mathcal N(\boldsymbol{\mu}_{\varphi}, \boldsymbol{\Sigma}_{\varphi})$ in the language space $\mathcal Y$ is derived from textual meta-knowledge $\mathbf s_u$, which may include noisy semantics that are irrelevant to the recommendation task \cite{ren2024representation}. Therefore, our goal is to match distributions between $\mathcal X$ and $\mathcal Y$ to mitigate the differences across spaces. Based on this, we propose three different matching strategies, which are elaborated in detail as follows.

\subsubsection{\textbf{Global Optimality for Distribution Matching}}

Given two distributions $q_{\phi}$ and $p_{\varphi}$, considering the structural differences between them, we first measure the transport cost using the Wasserstein distance \cite{xu2020learning, liu2022exploiting}, thereby capturing the geometric discrepancies across spaces:
\begin{equation}
\label{WD}
\text{D}_{n-\text{WD}}(p_\varphi, q_\phi) = \left [\min_{\gamma \in \prod(p_\varphi, q_\phi)}\int_{\mathcal X\times \mathcal Y}c(x,y)^n d\gamma (x,y)\right ]^{\frac{1}{n}}, 
\end{equation}
where $c:\mathcal X \times \mathcal Y \mapsto \mathbb R$ is a direct distance function between two spaces, $\gamma$ is the joint distribution between $q_{\phi}$ and $p_{\varphi}$, and $n$ refers to the order of the Wasserstein distance. Eq. \ref{WD} essentially involves finding an optimal transport plan between two distributions \cite{liu2022exploiting}, where the objective is to minimize the transport cost of matching one distribution with another. Given $p_{\varphi}\sim\mathcal N(\boldsymbol{\mu}_{\varphi}, \boldsymbol{\Sigma}_{\varphi})$ and $q_{\phi}\sim\mathcal N(\boldsymbol{\mu}_{\phi}, \boldsymbol{\Sigma}_{\phi})$, the 2-Wasserstein distance $\text{D}_{2-\text{WD}}(p_\varphi, q_\phi)$ is \cite{xu2020learning}:
\begin{equation}
\text{D}_{2-\text{WD}}(p_\varphi, q_\phi) = \left \|\ \boldsymbol{\mu}_{\varphi}-\boldsymbol{\mu}_{\phi}\right \|_2^2+\text{Tr}\left (\boldsymbol{\Sigma}_{\varphi} + \boldsymbol{\Sigma}_{\phi} - 2(\boldsymbol{\Sigma}_{\varphi}^{\frac{1}{2}}\boldsymbol{\Sigma}_{\phi}\boldsymbol{\Sigma}_{\varphi}^{\frac{1}{2}})^{\frac{1}{2} } \right ).
\end{equation}

For the recommendation task, we also consider optimizing the regularization term from the ELBO in Eq. \ref{elbo} to prevent the collaborative space $\mathcal X$ from being biased towards the language space $\mathcal Y$. Building on this, a dynamic trade-off process is established, combining the optimal transport matching term and the encoder's regularization term, leading to the proposed Global Optimality for Distribution Matching (GODM):
\begin{equation}
\label{GODM}
\mathcal L_{\text{GODM}}=\text{D}_{\text{KL}}(q_\phi||p_{\mathbf z}) + \beta \cdot \text{D}_{n-\text{WD}}(p_\varphi, q_\phi), 
\end{equation}
where $p_{\mathbf z}$ is the prior belief parameterized as standard Gaussian distribution $\mathcal N(\mathbf 0, \mathbf I)$, and $\beta$ here is a trade-off coefficient that controls the magnitude of the match. The matching term  $\mathcal L_{\text{GODM}}$ combines the advantages of optimal transport with the regularization effect of the KL divergence. This ensures that the distributions are not only matched in a geometrically optimal manner but also that the encoder $f_\phi$'s behavior is regularized, thereby enhancing generalization and maintaining consistency across spaces.

\subsubsection{\textbf{Composite Prior for Distribution Matching}}
In GODM, we measure the Wasserstein distance between two distributions to directly match them. A potential drawback is that when the input information contains excessive noise, the direct matching process may interfere with the optimization of both spaces, making the performance of GODM dependent on the quality of the input data. Considering that the reverse KL divergence is directly employed as a regularization term in the ELBO, we also explore whether the distribution $p_{\varphi}$ of the language space $\mathcal Y$ can be leveraged as prior knowledge for the distribution $q_{\phi}$ from the collaborative space $\mathcal X$, and the reverse KL divergence is treated as mode-seeking \cite{li2023mode}, guiding $q_{\phi}$ to match the orientations of distribution 
$p_{\varphi}$ that only place mass under appropriate constraints. Specifically, we introduce an intermediate composite prior $p_{\text{com}}$, which is a linear interpolation of two learned distributions:
\begin{equation}
\label{composite}
p_{\text{com}}=\alpha \cdot \mathcal N(\boldsymbol{\mu}_{\phi}, \boldsymbol{\Sigma}_{\phi}) + (1 - \alpha) \cdot \mathcal N(\boldsymbol{\mu}_{\varphi}, \boldsymbol{\Sigma}_{\varphi}), 
\end{equation}
where $\alpha \in [0, 1]$ is a weighting coefficient (set to 0.5 by default). Subsequently, we use the composite prior $p_{\text{com}}$ as a bridge to separately quantify the distributional differences between the variational distribution $q_{\phi}$ and the approximate posterior distribution $p_{\varphi}$:
\begin{equation}
\label{JS_divergence}
\text{D}_{\text{CP}}(p_\varphi, q_\phi) = \text{D}_{\text{KL}}(p_\varphi||p_{\text{com}}) + \text{D}_{\text{KL}}(q_\phi || p_{\text{com}}),
\end{equation}
where $p_{\text{com}}$ is the composite prior derived from Eq. \ref{composite}. Since Eq. \ref{JS_divergence} involves the composite distribution $p_{\text{com}}$ from $q_{\phi}$ and $p_{\varphi}$, it exhibits greater tolerance to minor variations between distributions and is less sensitive to events with zero probability. Furthermore, Eq. \ref{JS_divergence} imposes additional constraints on the guiding capability of the language space $\mathcal Y$, ensuring that the distribution partially relies on the variational distribution $q_{\phi}$, thereby avoiding excessive concentration or divergence. When $\alpha=0.5$, the Eq. \ref{JS_divergence} essentially becomes the standard Jensen-Shannon (JS) divergence \cite{englesson2021generalized}. Building on this, we combine this matching term with the previously optimized encoder's KL regularization to propose the Composite Prior for Distribution Matching (CPDM):
\begin{equation}
\label{CPDM}
\mathcal L_{\text{CPDM}}=\text{D}_{\text{KL}}(q_\phi||p_{\mathbf z}) + \beta \cdot \text{D}_{\text{CP}}(p_\varphi, q_\phi),
\end{equation}
where $\beta$ is also a trade-off coefficient here, and its definition is analogous to that in Eq. \ref{GODM}.

\subsubsection{\textbf{Mixing Divergence for Distribution Matching}}
In the previous sections, GODM is a strategy that directly matches the geometric shapes of two distributions, which may excessively interfere with the original distribution. In contrast, CPDM adopts an indirect guiding process that models the collaborative space by composite prior $p_{\text{com}}$, but it introduces additional computational overhead.
 Therefore, a new question arises: \textit{Can we simplify the design of CPDM and achieve the matching process of GODM in an indirect manner?} A feasible approach is to directly set the distribution $p_\varphi$ of the language space $\mathcal Y$ as the prior for the collaborative space $\mathcal X$, as follows \cite{kingma2013auto, zhang2024dual}:
\begin{equation}
\begin{aligned}
\label{KL}
\text{D}_{\text{KL}}(q_\phi||p_{\varphi})&=\frac{1}{2} [( \text{Tr}\left( \boldsymbol\Sigma_\varphi^{-1} \boldsymbol\Sigma_\phi \right) \\ &+ (\boldsymbol\mu_\varphi - \boldsymbol\mu_\phi)^\top \boldsymbol\Sigma_\varphi^{-1} (\boldsymbol\mu_\varphi - \boldsymbol\mu_\phi) - d + \log \frac{\det \boldsymbol\Sigma_\varphi}{\det \boldsymbol\Sigma_\phi} ],
\end{aligned}
\end{equation}
where $d$ is the dimension of the input distributions. Similar to GODM, such direct optimization will cause the variational distribution $q_\phi$ learned in the collaborative space $\mathcal X$ to gradually resemble the approximate distribution $p_\varphi$ from the language space $\mathcal Y$ \cite{salah2021towards}, thereby causing the original semantic information of the collaborative space $\mathcal X$ to be progressively lost \cite{chen2024hate}. Therefore, we propose the Mixing Divergence for Distribution Matching (MDDM), aiming to reconcile the KL regularization term from the ELBO in Eq. \ref{elbo} with the KL matching term presented in Eq. \ref{KL}:
\begin{equation}
\label{MDDM}
\mathcal L_{\text{MDDM}}=\beta \cdot\text{D}_{\text{KL}}(q_\phi||p_{\mathbf z}) + (1- \beta) \cdot \text{D}_{\text{KL}}(q_\phi||p_{\varphi}).
\end{equation}
Please note that $\beta \in [0,1]$ is considered here as a mixing coefficient, which can be manually adjusted or controlled via distribution sampling \cite{zhang2018mixup}. Compared to the previously proposed GODM and CPDM, the MDDM presented in this section is more concise and directly integrates with the ELBO objective, without the need for additional intermediate terms. And for the KL divergence $\text{D}_{\text{KL}}(q_\phi||p_{\varphi})$, according to \cite{salah2021towards},  we provide the following analysis:

\begin{align}
\label{mi}
\text{D}_{\text{KL}}(q_\phi||p_{\varphi})&=\mathbb E_{q(\mathbf x_u, \mathbf s_u)q_\phi(\mathbf z_u|\mathbf x_u)}\left [ \text{log}\frac{q_\phi(\mathbf z_u|\mathbf x_u)}{p_\varphi(\mathbf z_u|\mathbf s_u)} \right ] \nonumber\\
&=\mathbb E_{q(\mathbf x_u, \mathbf s_u)q_\phi(\mathbf z_u|\mathbf x_u)}\left [ \text{log}\frac{q_\phi(\mathbf z_u,\mathbf x_u)}{q(\mathbf x_u)q_{\phi}(\mathbf z_u)}\times \frac{q_{\phi}(\mathbf z_u)}{p_\varphi(\mathbf z_u|\mathbf s_u)} \right ] \nonumber\\
&=\mathbb I(\mathbf z_u; \mathbf x_u) - \mathbb E_{q(\mathbf x_u, \mathbf s_u)q_\phi(\mathbf z_u|\mathbf x_u)}\left [ \text{log}\frac{p_\varphi(\mathbf z_u,\mathbf s_u)}{p(\mathbf s_u)q_{\phi}(\mathbf z_u)} \right ] \nonumber\\
&= \mathbb I(\mathbf z_u; \mathbf x_u) - \tilde{\mathbb I} (\mathbf z_u; \mathbf s_u),
\end{align}
where $\mathbb I(\mathbf z_u; \mathbf x_u)$ is the mutual information between $\mathbf z_u$ and $\mathbf x_u$. Since the direct calculation of mutual information $\mathbb I(\mathbf z_u; \mathbf s_u)$ is challenging, a variational lower bound $\tilde{\mathbb I} (\mathbf z_u; \mathbf s_u)$ is used for approximation, as $\mathbb E_{q_\phi}[\text{log}q_{\phi}]\ge \mathbb E_{q_\phi}[\text{log}p_{\varphi}]$ holds true due to the non-negativity of the KL divergence \cite{luo2022understanding}. Intuitively, Eq. \ref{mi}
decomposes the reverse KL divergence into the difference between two mutual information terms. According to the information bottleneck theory \cite{cao2022cross}, the following conclusion holds \cite{salah2021towards}:
\begin{align}
\label{ib}
\mathcal L_{\text{IB}}&=\mathbb I(\mathbf z_u; \mathbf s_u) - \beta \cdot \mathbb I(\mathbf z_u; \mathbf x_u) \nonumber\\
&\ge \tilde{\mathbb I} (\mathbf z_u; \mathbf s_u) - \mathbb I(\mathbf z_u; \mathbf x_u)=-\text{D}_{\text{KL}}(q_\phi||p_{\varphi}),
\end{align}
where $\beta \in [0, 1]$ is considered here as a Lagrangian multiplier for adaptive trade-off. This establishes the relationship between KL divergence and information bottleneck, and transforms the objective into an optimizable form. By optimizing the variational distribution $q_\phi$ and the approximate posterior $p_\varphi$, we can achieve a trade-off between information compression and generation quality \cite{salah2021towards, truong2021bilateral}. Furthermore, MDDM is more streamlined than CPDM, as it does not require constructing additional priors and can simultaneously coordinate both the regularization and matching terms.
% Specifically, optimizing the mutual information ensures that $\mathbf z_u$ retains more information from the language space $\mathcal Y$ while reducing redundant information from the collaborative space $\mathcal X$. This makes the model's predictions more dependent on the language space $\mathcal Y$, which is particularly beneficial for recommendation tasks with sparse interactions. A similar inference can be made for the proposed CPDM, as the KL divergence has an analytical solution, and the additivity property of Gaussian distributions facilitates straightforward computation \cite{luo2022understanding, wang2023diffusion}.

In addition, considering that the language space $\mathcal Y$ may contain noise unrelated to the recommendation task \cite{ren2024representation}, it is crucial to carefully evaluate the guidance provided by the language space $\mathcal Y$. Therefore, in the proposed MDDM, we introduce the concept of mixing divergence, which aims to impose constraints on the structural matching of the probability spaces. The mixing design simultaneously scales the regularization term $\text{D}_{\text{KL}}(q_\phi||p_{\mathbf z})$ from the ELBO, which is similar in spirit to works like $\beta$-VAE \cite{higgins2017beta}. This strategy is designed to adjust the generative model in a way that encourages the learning of more disentangled representations, while preserving reconstruction information to the greatest extent possible \cite{liang2018variational}.

\begin{algorithm}
\caption{The training process of \textsf{DMRec}}
\label{algorithms}
\KwIn{user–item ineraction $\mathbf X$, prompt template $\mathcal P$; language model $f_{\text{LM}}$, text embedding model $f_{\text{text}}$, base generative model $f_{\phi, \theta}$, and meta network $f_\varphi$.}
\begin{algorithmic}[1]
\STATE initialize parameters for $f_{\phi, \theta}$ and $f_\varphi$;
\STATE retrieve all user/item profiles by $f_{\text{LM}}$ with  $\mathcal P$ (Eq . \ref{prompt});
\WHILE {\textsf{DMRec} not converge}
\STATE sample a mini-batch of user set $\mathcal O$;
\FOR{$u\in \mathcal O$}
\STATE retrieve the semantic embeddings of user $u$ and interacted items $\mathcal M(u)$ by $f_{\text{text}}$ (Eq . \ref{text_embedding});
\STATE calculate the variational distribution $q_{\phi}$ in the collaborative space $\mathcal X$ by $f_{\phi}$ (Eq. \ref{vae_q});
\STATE calculate the approximate distribution $p_{\varphi}$ in the language space $\mathcal Y$ by $f_\varphi$ (Eq. \ref{language_p});
\STATE reconstruct whole interactions for user $u$ by $f_{\theta}$;
\STATE calculate the recommendation loss $\mathcal L_{\text{rec}}$ by Eq. \ref{reconstruction};
\STATE calculate the matching loss $\mathcal L_{\text{DM}}$ by GODM (Eq. \ref{GODM}), CPDM (Eq. \ref{CPDM}), or MDDM (Eq. \ref{MDDM});
\ENDFOR
\STATE average gradients from mini-batch;
\STATE update parameter by descending the gradients $\nabla_{\phi, \varphi, \theta}\mathcal L$;

\ENDWHILE
\RETURN model parameters $\phi, \varphi, \theta$;
\end{algorithmic}
\end{algorithm}

\subsection{DMRec Framework}

In the previous section, we introduce three strategies for cross-space distribution matching. Based on this, we propose the Distribution Matching-based Framework for Generative Recommendation (\textsf{DMRec}), which can serve as a plug-and-play component for probabilistic generative recommendation models.

Given a generative model that includes an encoder parameterized by $\phi$, with the input being historical interactions $\mathbf x_u$, the user's approximate distribution $q_{\phi}(\mathbf z_u|\mathbf x_u)$ can be obtained through the encoder. Subsequently, the decoder parameterized by $\theta$ is responsible for reconstructing the complete interactions $p_{\theta}(\mathbf x_u|\mathbf z_u) $ based on the variational distribution $\mathbf z_u$. According to Eq. \ref{elbo}, the generative model is optimized via the negative reconstruction error \cite{liang2018variational}:
\begin{equation}
\label{reconstruction}
\mathcal L_{\text{rec}}= \mathbb E_{q_{\phi}(\mathbf z_u|\mathbf x_u)q_{\varphi}(\mathbf z_u|\mathbf s_u)}[\text{log}p_{\theta}(\mathbf x_u|\mathbf z_u^*)].
\end{equation}
As revisited in Section 2.3, we construct the user's approximate distribution $p_{\varphi}(\mathbf z_u|\mathbf s_u)$ in the language space $\mathcal Y$ parameterized by $\varphi$. And in Section 2.4, we propose three distinct strategies for distribution matching between the language space $\mathcal Y$ and the collaborative space $\mathcal X$. Therefore, to guide both the recommendation task and the matching task, \textsf{DMRec} adopts a multi-task learning strategy to jointly optimize these parameters:
\begin{equation}
\mathcal L_{\text{\textsf{DMRec}}}= \mathcal L_{\text{rec}} + \mathcal L_{\text{DM}},
\end{equation}
where the matching loss $\mathcal L_{\text{DM}}$ can be one of the strategies proposed in Section 2.4: GODM (Eq. \ref{GODM}), CPDM (Eq. \ref{CPDM}), or MDDM (Eq. \ref{MDDM}), and already includes the regularization term of the original ELBO. The training process of \textsf{DMRec} is shown in Algorithm \ref{algorithms}.

\section{Experiments}
In this section, we conduct extensive experimental analysis on three widely used datasets to validate the effectiveness of \textsf{DMRec}.
\subsection{Experiment Settings}
\subsubsection{\textbf{Datasets}} To conduct experimental analysis, we adopt three widely used recommendation datasets: Amazon-Book, Yelp, and Steam \cite{ren2024representation}, which are varied in scale, field, and sparsity. Table summarizes the scales of all datasets after pre-processing. In line with existing studies \cite{liang2018variational, he2020lightgcn}, all datasets employ implicit feedback, with interactions rated below 3 being excluded \cite{ren2024representation}. The datasets are partitioned into training, validation, and test sets at a ratio of 3:1:1. The statistical information is shown in Table \ref{dataset}.

\begin{table}[t]
\small
\setlength{\abovecaptionskip}{0.1cm}
\setlength{\belowcaptionskip}{0.1cm} 
  \caption{ Statistics of the datasets.}
  \label{dataset}
  \begin{tabular}{l|c|c|c|c}
    \hline
    \textbf{Dataset}&\textbf{\#Users}&\textbf{\#Items}&\textbf{\#Interactions}&\textbf{Sparsity}\\
    \hline
    \hline
    \textbf{Amazon-Book}&11,000&9,332&200,860&99.80\%\\
    \textbf{Yelp}&11,091&11,010&277,535&99.77\%\\
    \textbf{Steam}&23,310&5,237&525,922&99.57\%\\
    \hline
  \end{tabular}
\end{table}

\begin{table*}
\small
        \centering
\setlength{\abovecaptionskip}{0.1cm}
\setlength{\belowcaptionskip}{0.1cm} 
  \caption{Overall performance comparisons between \textsf{DMRec} and base models on Amazon-Book, Yelp, and Steam datasets \textit{w.r.t.} Recall@N and NDCG@N (N $\in [10, 20]$). The best-performing model is highlighted in \textbf{bold}, whereas the second-best model is shown in \underline{underlined}. Improv.\% refers to the relative improvement of the best-performing model compared to the base model.}
  \label{performance1}
  \begin{tabular}{c|c|cccc|cccc|cccc}
    \hline
    \multicolumn{2}{c|}{\textbf{Model}}&\multicolumn{4}{c|}{\textbf{Amazon-Book}}&
    \multicolumn{4}{c|}{\textbf{Yelp}}&
    \multicolumn{4}{c}{\textbf{Steam}}\\
	\cline{1-14}		Base model&Variants&R@10&R@20&N@10&N@20&R@10&R@20&N@10&N@20&R@10&R@20&N@10&N@20\\
    \hline
    \hline
    \multirow{6}{*}{Mult-VAE} &Base \cite{liang2018variational} & 0.1013 & 0.1498 & 0.0771 & 0.0936 & 0.0732 & 0.1189 & 0.0593 & 0.0751 & 0.0929 &0.1452 & 0.0744 & 0.0923\\
    \cline{2-14}
    ~ &\textsf{DMRec-G} & \underline{0.1047} & 0.1545 & 0.0795 & 0.0960 & 0.0752 & 0.1226 & 0.0612 &0.0775 & \underline{0.0954} & 0.1495 & \underline{0.0764} & \underline{0.0950}\\
    ~ & \textsf{DMRec-C} & \underline{0.1047} & \underline{0.1561} & \underline{0.0802} & \underline{0.0971} & \textbf{0.0771} &\underline{0.1254} & \textbf{0.0625} & \textbf{0.0792} &0.0952 & \underline{0.1496} & 0.0762 & 0.0948\\
    ~ & \textsf{DMRec-M} & \textbf{0.1069} & \textbf{0.1571} & \textbf{0.0812} & \textbf{0.0979} & 0.0768 & \textbf{0.1261} & 0.0618 & 0.0785 & \textbf{0.0986} & \textbf{0.1536} & \textbf{0.0784} & \textbf{0.0973}\\
    \cline{2-14}
     ~ & Improv.\% & 5.53\% & 4.87\% & 5.32\% & 4.59\% & 5.33\% & 6.06\% & 5.40\% & 5.46\% & 6.14\% & 5.79\% & 5.38\% & 5.42\%\\
~ & $p$-values & 5.82e-7 & 5.62e-8 & 2.62e-6 & 6.67e-7 & 8.52e-4 & 3.03e-7 & 6.88e-5 & 9.44e-7 & 6.24e-9 & 1.45e-9 & 8.17e-9 & 7.19e-10\\
      \hline
        \multirow{6}{*}{L-DiffRec} &Base \cite{wang2023diffusion} & 0.1048 & 0.1495 & 0.0844 & 0.0990 & 0.0721 & 0.1177 & 0.0590 & 0.0745 &0.0885 & 0.1395 & 0.0722 & 0.0893\\
    \cline{2-14}
    ~ &\textsf{DMRec-G} & \underline{0.1063} & \underline{0.1529} & \underline{0.0867} & \underline{0.1016} & 0.0750 & \underline{0.1225} & \underline{0.0625} & \underline{0.0787} &0.0899 & 0.1419 & 0.0740 & 0.0916\\
    ~ & \textsf{DMRec-C} & 0.1059 & 0.1525 & 0.0856 & 0.1006 & \textbf{0.0766} & \textbf{0.1248} & \textbf{0.0641} & \textbf{0.0804} & \underline{0.0907} & \underline{0.1420} & \underline{0.0744} & \underline{0.0917}\\
    ~ &\textsf{DMRec-M}& \textbf{0.1111} & \textbf{0.1576} & \textbf{0.0893} & \textbf{0.1044} & \underline{0.0753} & 0.1218 & 0.0613 & 0.0771 & \textbf{0.0961} &\textbf{0.1474} &\textbf{0.0773} &\textbf{0.0949}\\
    \cline{2-14}
     ~ & Improv.\% & 6.01\% & 5.42\% & 5.81\% & 5.46\% & 6.24\% & 6.03\% & 8.64\% & 7.92\% & 8.59\% & 5.66\% & 7.06\% & 6.27\%\\
     ~ & $p$-values & 8.67e-4 & 4.19e-4 & 8.75e-4 & 1.13e-5 & 5.33e-7 & 1.46e-7 & 1.07e-7 & 3.11e-8 & 2.27e-5 & 3.61e-6 & 2.13e-5 & 3.06e-6\\
         \hline
    \multirow{6}{*}{CVGA} &Base \cite{zhang2023revisiting} & 0.1030 & 0.1522 & 0.0799 & 0.0960 & 0.0779 & 0.1249 & 0.0639 & 0.0801 &0.0842 &0.1339 & 0.0679 & 0.0847\\
    \cline{2-14}
    ~ &\textsf{DMRec-G} & \textbf{0.1135} & \underline{0.1647} & \underline{0.0865} & 0.1035 & 0.0806 & \textbf{0.1310} & \textbf{0.0657} & \textbf{0.0829} & 0.0959 & 0.1506 & 0.0771 & 0.0958\\
    ~ &\textsf{DMRec-C} & \underline{0.1132} & 0.1642 & \textbf{0.0869} & \underline{0.1038} & \textbf{0.0809} & \underline{0.1307} &\underline{0.0655} &\underline{0.0826} &\underline{0.0973} & \underline{0.1522} & \underline{0.0781} & \underline{0.0969}\\
    ~ &\textsf{DMRec-M} & 0.1129 & \textbf{0.1652} & \textbf{0.0869} & \textbf{0.1042} & 0.0803 & 0.1304 & 0.0654 & \underline{0.0826} & \textbf{0.0989} & \textbf{0.1536} & \textbf{0.0792} & \textbf{0.0979}\\
    \cline{2-14}
     ~ & Improv.\% & 10.19\% & 8.54\% & 8.76\% & 8.54\% & 3.85\% & 4.88\% & 2.82\% & 3.50\% & 17.46\% & 14.71\% & 16.64\% & 15.58\%\\
     ~ & $p$-values & 1.39e-11 & 3.7e-12 & 9.2e-12 & 6.01e-12 & 3.41e-4 & 3.61e-8 & 3.89e-4 & 7.20e-6 & 1.94e-12 & 2.99e-12 & 7.04e-13 & 3.3e-13\\

\hline
  \end{tabular}
\end{table*}

\subsubsection{\textbf{Base Models}}
\label{base_model}
For the proposed \textsf{DMRec}, we select the following representative generative models as the base models:
\begin{itemize}[leftmargin=*]
    \item \textbf{Mult-VAE} \cite{liang2018variational} is a classic generative recommendation method based on the vanilla VAE with multinomial likelihood.
    \item \textbf{CVGA} \cite{zhang2023revisiting} extends the basic VAE by incorporating graph structure modeling, with its overall design being similar to that of a variational graph auto-encoder.
    \item \textbf{L-DiffRec} \cite{wang2023diffusion} attempts to introduce diffusion models into generative recommendation, adding Gaussian noise progressively in the distribution space rather than directly to the original data.
\end{itemize}
For \textsf{DMRec}, the models using GODM, CPDM, and MDDM are abbreviated as \textbf{\textsf{DMRec-G}}, \textbf{\textsf{DMRec-C}}, and \textbf{\textsf{DMRec-M}}, respectively.

\subsubsection{\textbf{LM-enhanced Baselines}}
For a more comprehensive comparison, we also consider selecting the following LM-enhanced works as baselines:
\begin{itemize}[leftmargin=*]
    \item \textbf{KAR} \cite{xi2024towards} creates textual profiles for users and items, and integrates the LM-enhanced representations with recommenders through a hybrid-expert adaptor.
    \item \textbf{LLMRec} \cite{wei2024llmrec} enhances data reliability by employing graph augmentation strategies based on language models and a denoising mechanism for data robustification.
    \item \textbf{RLMRec} \cite{ren2024representation} aligns semantic representations of users and items between the LM-enhanced and recommendation representations. The contrastive strategy is referred to as RLMRec-C, while the generative strategy is referred to as RLMRec-G.
    \item \textbf{AlphaRec} \cite{sheng2024language} is a recently proposed LM-based recommendation method that directly applies non-linear mapping and graph convolution operations on LM-enhanced item representations.
    \item \textbf{AlphaRec*} is a variant of AlphaRec \cite{sheng2024language} additionally utilizes user-side embedding representations learned from language model.
\end{itemize}
The comparative methods we selected exclude fine-tuning-based methods (\textit{e.g.}, TallRec \cite{bao2023tallrec} and LLaRA \cite{liao2024llara}). Specifically, their research is orthogonal to ours as they primarily focus on fine-tuning LM and are mainly applied to sequential recommendation \cite{kang2018self}.

\subsubsection{\textbf{Implementation Details}} We implement \textsf{DMRec} by PyTorch \footnote{https://github.com/BlueGhostYi/DMRec}. All models are initialized by Xavier \cite{glorot2010understanding} and optimized by Adam \cite{kingma2014adam}. The structure of all baseline models follows the default settings from the original papers. The default batch size is 1,024. For all LM-enhanced methods, we use OpenAI's GPT-3.5-turbo as the language model and text-embedding-ada-002 \cite{neelakantan2022text} for text embeddings to ensure a fair comparison. The design details of the prompt $\mathcal P$ are outlined in \cite{ren2024representation}. To evaluate the recommendation performance, we use the metrics Recall@N and NDCG@N \cite{he2017neural}. For each user, the full-ranking strategy \cite{he2020lightgcn} is employed. Early stopping is triggered if Recall@20 on the validation set fails to improve for 20 consecutive iterations. To mitigate bias, we run the experiments with 10 different random seeds and report the average results.

\subsection{Performance Comparisons}
\subsubsection{\textbf{Model-agnostic Performance Comparison}}
\begin{figure}
\setlength{\abovecaptionskip}{0.0cm}
\setlength{\belowcaptionskip}{0.0cm} 
\centering
\subfigure[Amazon-Book]{\includegraphics[width=0.15\textwidth]{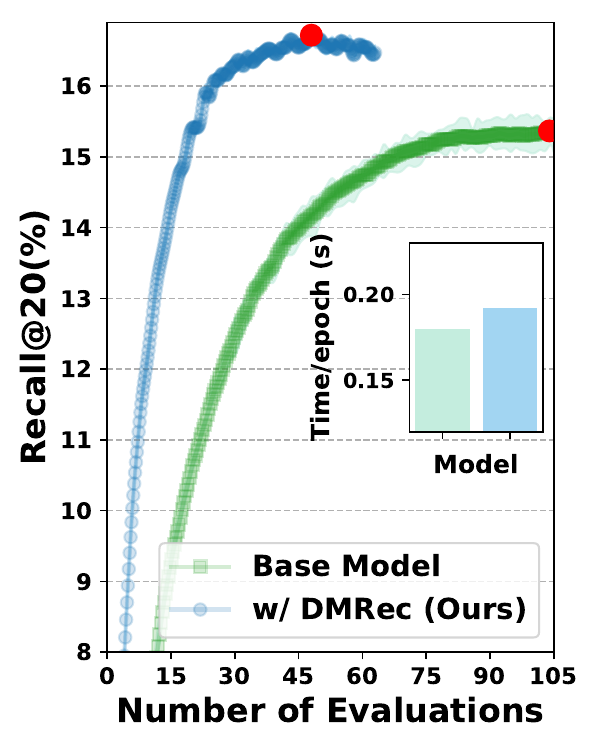}
\label{speed_amazon_case}}
\hfil
\subfigure[Yelp]{\includegraphics[width=0.15\textwidth]{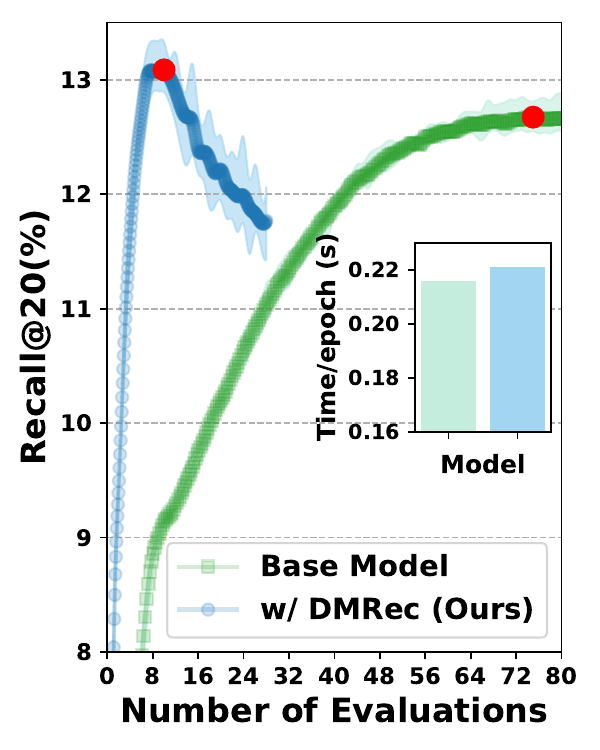}
\label{speed_yelp_case}}
\hfil
\subfigure[Steam]{\includegraphics[width=0.15\textwidth]{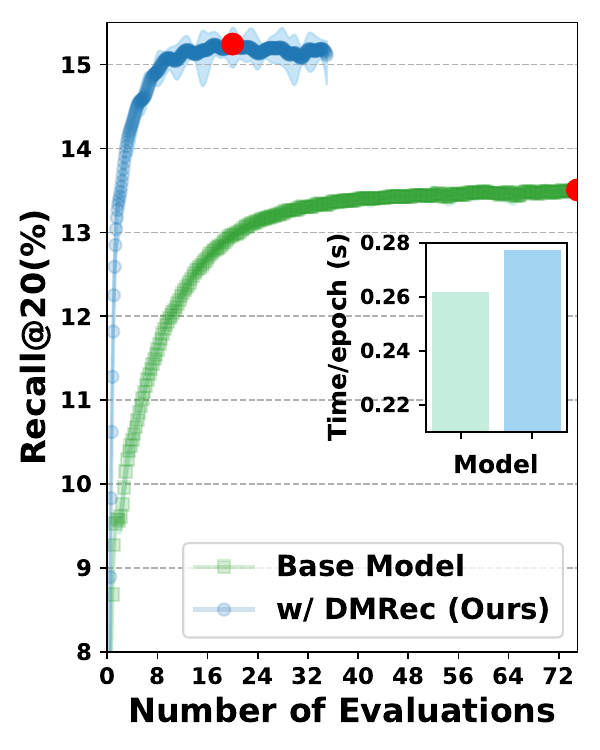}
\label{speed_steam_case}}

\caption{Comparison of the training process and speed of the base model and \textsf{DMRec} \textit{w.r.t.} Recall@20 on validation sets. The red dot indicates the best-performing on test sets.}
\label{fig_speed}
\end{figure}

To verify the generalization ability of \textsf{DMRec}, we apply it to three basic generative models listed in Section \ref{base_model}. The experimental results are shown in Table \ref{performance1}, leading to the following findings:
\begin{itemize}[leftmargin=*]
    \item Intuitively, after incorporating the three different matching strategies of \textsf{DMRec}, the recommendation performance of the basic models shows varying degrees of improvement across the three datasets. Taking Mult-VAE with the incorporation of MDDM as an example, \textsf{DMRec-M} improves by 4.87\%, 6.06\% and 5.79\% over the base model \textit{w.r.t.} Recall@20 on Amazon-Book, Yelp, and Steam datasets, respectively. The above experimental results demonstrate the generalizability of the proposed \textsf{DMRec}.
    \item Overall, GODM achieves the best performance only on Yelp dataset. And CPDM achieves the second-best performance while MDDM is the best-performing model in most cases. This is because GODM heavily depends on the quality of the input data, and MDDM can collaboratively optimize both regularization and matching terms for a more fine-grained trade-off.
    \item It is worth noting that although CVGA performs poorly on Steam dataset, the incorporation of \textsf{DMRec} allows it to regain its competitiveness  in this scenario. The substantial performance improvement further demonstrates the effectiveness of \textsf{DMRec}.
    \item Finally, we take CVGA as an example to compare the training efficiency between the base model and \textsf{DMRec}, as shown in Fig. \ref{fig_speed}. \textsf{DMRec} significantly improves the training efficiency of the base model, while the time per iteration only increases slightly.
\end{itemize}

\subsubsection{\textbf{Performance Comparison with LM-enhanced Methods}}
Going further, to verify the effectiveness of \textsf{DMRec}, we compare it with other LM-enhanced recommendation methods. We select Mult-VAE \cite{liang2018variational} as the base model for \textsf{DMRec} with the MDDM strategy, and LightGCN \cite{he2020lightgcn} for LM-enhanced baselines due to structural differences. The experimental results are presented in Table \ref{tab:LM}. 
Intuitively, \textsf{DMRec} achieves the best recommendation performance on both metrics across all datasets. This can be attributed to two main factors: on the one hand, generative methods aim to generate probabilities for users across the entire set of items, and this estimation process does not rely on pairwise embedding modeling, thereby reducing the excessive dependence on sparse interactions. On the other hand, cross-space distribution matching provides a feasible course of action for guiding the recommendation process with language models, achieving adaptive trade-offs. This process not only coordinates the optimization process but also helps to avoid issues such as posterior collapse as much as possible.

\begin{table}[!t]

\setlength{\abovecaptionskip}{0.1cm}
\setlength{\belowcaptionskip}{0.1cm} 
\centering
\small
  \caption{Comparison of recommendation performance between \textsf{DMRec} and LM-enhanced recommendation methods on three datasets \textit{w.r.t.} Recall@20 and NDCG@20.}
  \label{tab:LM}
  \begin{tabular}{l|cc|cc|cc}
    \hline
        &\multicolumn{2}{c|}{\textbf{Amazon-Book}}&
    \multicolumn{2}{c|}{\textbf{Yelp}}&\multicolumn{2}{c}{\textbf{Steam}}\\
	\cline{2-7}	
    &R@20&N@20&R@20&N@20&R@20&N@20\\
    \hline
    \hline
KAR&0.1416&0.0863&0.1194&0.0756&0.1353& 0.0854\\
    \hline
LLMRec&0.1469&0.0855&0.1203&0.0751&0.1431&0.0901\\
\hline
RLMRec-C&\underline{0.1483}&\underline{0.0903}&\underline{0.1230}&\underline{0.0776}&0.1421&0.0902\\
RLMRec-G&0.1446&0.0887&0.1209&0.0761&\underline{0.1433}&\underline{0.0907}\\
    \hline
AlphaRec&0.1412&0.0873&0.1212&0.0752&0.1404&0.0889\\
AlphaRec*&0.1421&0.0835&0.1213&0.0752&0.1420&0.0898\\
\hline
\hline
\textbf{\textsf{DMRec-M}}&\textbf{0.1571}&\textbf{0.0979}&\textbf{0.1261}&\textbf{0.0785}&\textbf{0.1536}&\textbf{0.0973}\\
   \hline
\end{tabular}

\end{table}

\subsection{In-depth Analysis of DMRec}

\subsubsection{\textbf{Performance Comparison w.r.t. Data Sparsity}}

\begin{figure}
\setlength{\abovecaptionskip}{0.0cm}
\setlength{\belowcaptionskip}{0.0cm} 
\centering
\subfigure[Yelp]{\includegraphics[width=1.62in]{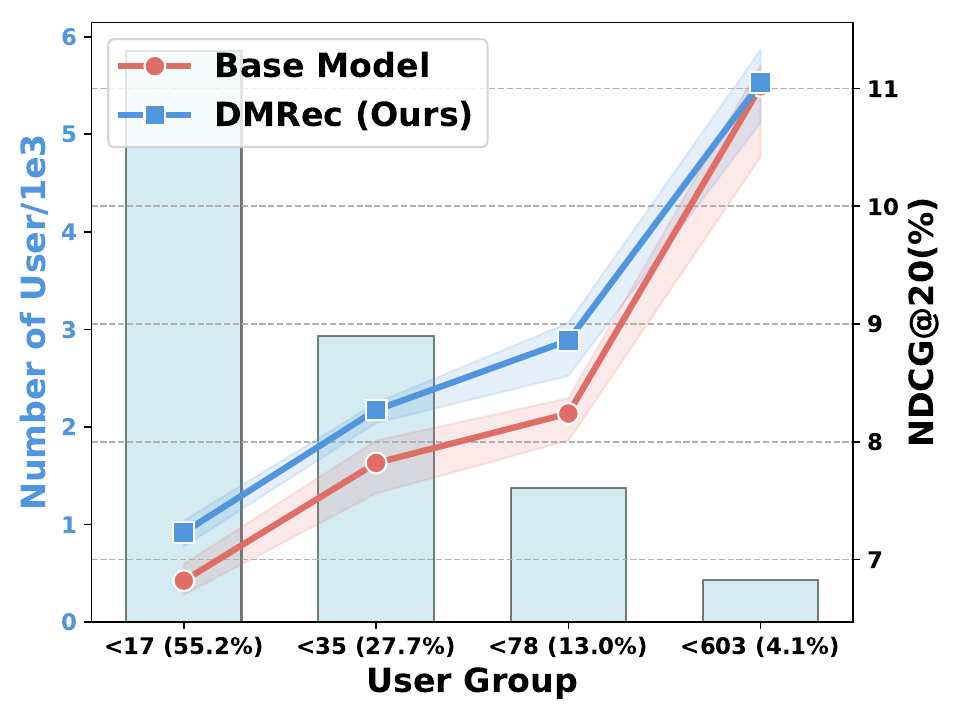}
\label{yelp_sparsity_case}}
\hfil
\subfigure[Steam]{\includegraphics[width=1.62in]{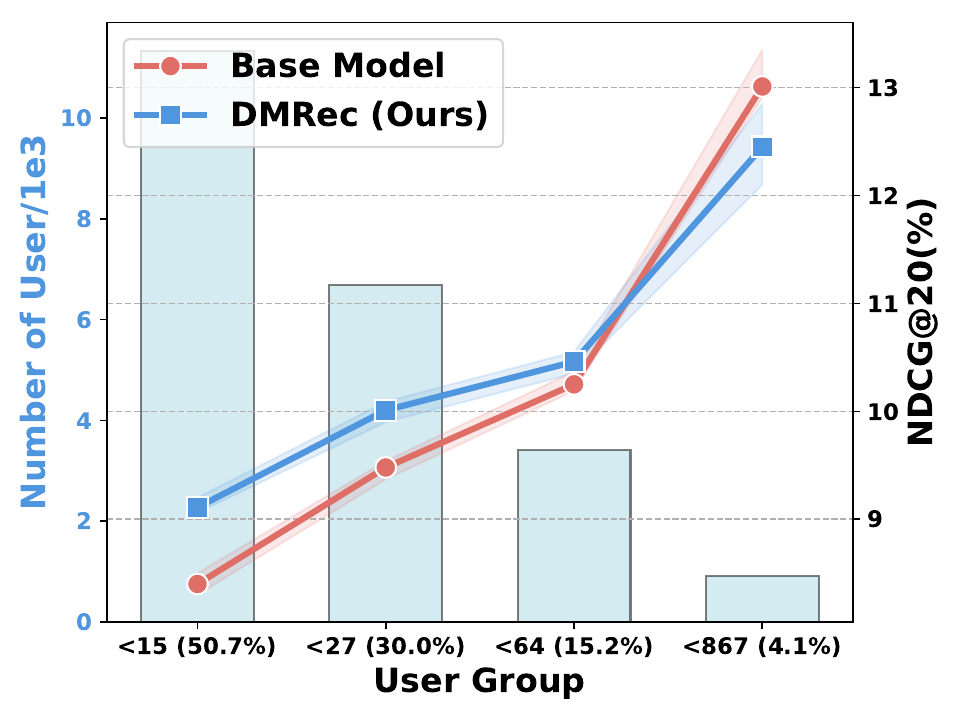}
\label{steam_sparsity_case}}

\caption{Comparison of the base model and \textsf{DMRec} performance across sparse user groups. The bar chart (left $y$-axis) shows user count, the line chart (right $y$-axis) shows NDCG@20, and the $x$-axis displays the number of interactions and user group proportions.}
\label{fig_sparsity}
\end{figure}

The sparsity issue has always been a core factor limiting recommendation performance \cite{yu2022graph}. To investigate whether DMRec can alleviate this challenge, we conduct a sparsity test on both Mult-VAE and \textsf{DMRec} based on MDDM. Specifically, following the strategy in \cite{zhang2023revisiting, yu2022graph}, we divide users into four groups according to the interaction scales and test each group separately. The experimental results are shown in Fig. \ref{fig_sparsity}. Intuitively, the proportion of users in the sparsest group is very high (exceeding 50\% in both datasets), indicating the extremely sparse nature of these recommendation scenarios. \textsf{DMRec} achieves significant improvements in the sparsest group while maintaining comparable performance to the base model in the dense user group. This suggests that \textsf{DMRec} places more emphasis on preference prediction for sparse users. We attribute this change and improvement to the benefits brought by the language model. Specifically, the user profiles generated by the language model can extract implicit preferences from extremely limited interaction data, thereby reconstructing more enriched user interactions.

\subsubsection{\textbf{Comparison w.r.t. Distribution Activity}} 
\begin{figure}
\setlength{\abovecaptionskip}{0.0cm}
\setlength{\belowcaptionskip}{0.0cm} 
\centering
\subfigure[Amazon-Book]{\includegraphics[width=0.15\textwidth]{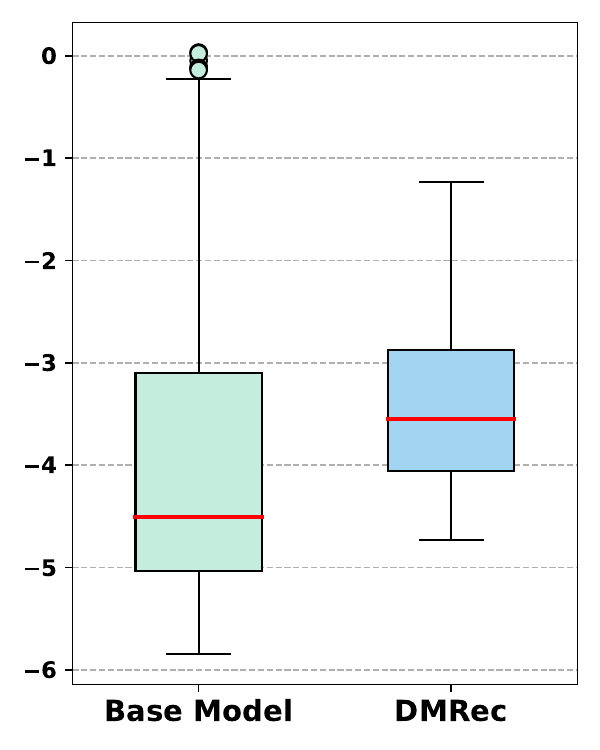}
\label{intent_case}}
\hfil
\subfigure[Yelp]{\includegraphics[width=0.15\textwidth]{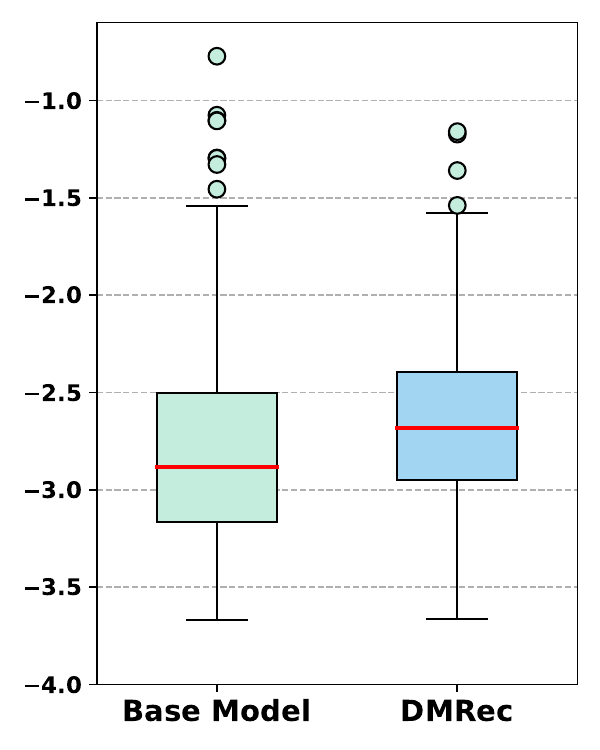}
\label{yelp_box_case}}
\hfil
\subfigure[Steam]{\includegraphics[width=0.15\textwidth]{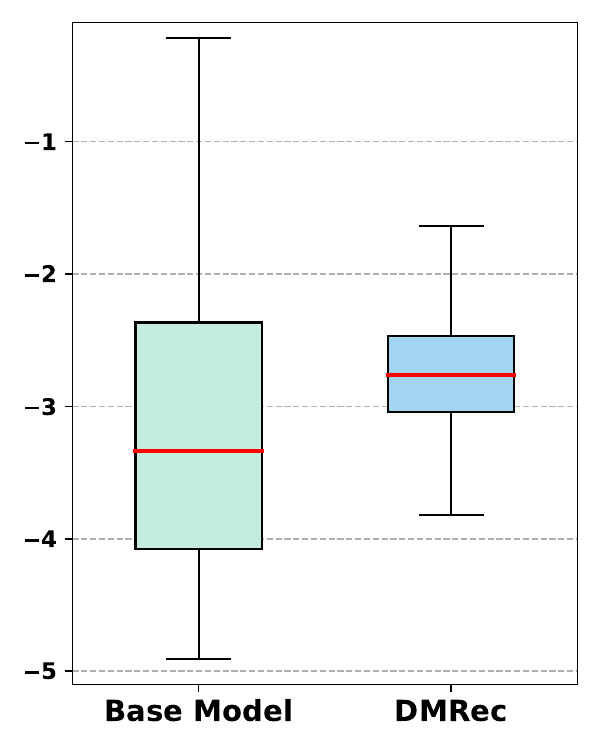}
\label{steam_box_case}}

\caption{Comparison of the base model and \textsf{DMrec} for user distribution dimension activity $\text{log} a^{\phi}$ on three datasets.}
\label{fig_box}
\end{figure}

Generative models often face a trade-off between representation ability and tractability. Simple model structures typically result in weaker expressiveness and posterior collapse \cite{kingma2016improved}. Therefore, following \cite{burda2015importance}, we use the metric $\text{log} a^{\phi}$ to measure the variation in each dimension of the distribution $q_\phi$, where $a_k^\phi=\text{Cov}_{p(\mathbf x)}(\mathbb E_{q_\phi(\mathbf z|\mathbf x)}[z_k])$. Given any dimension $k$, if the distribution $z_k$ encodes useful information, the expectation will exhibit variations across different users, thereby indicating its activity \cite{truong2021bilateral}. Therefore, we measure the distribution activities for the base model (based on Mult-VAE) and \textsf{DMRec} (based on MDDM) on three datasets, with the results shown in Fig. \ref{fig_box}. After introducing \textsf{DMRec}, the variance of the distribution on three datasets overall exhibits higher box spans and medians, while maintaining fewer outliers. Compared to the base model, \textsf{DMRec} has a larger variance and a relatively stable range, indicating that each distribution dimension experiences significant changes for different users. This allows it to encode more useful information without altering the model structure or the number of parameters \cite{truong2021bilateral}. Furthermore, due to the greater variability across different dimensions, the model is less likely to fall into the collapse trap \cite{higgins2017beta, wang2022posterior}, thus avoiding additional adjustment costs.

\subsubsection{\textbf{Ablation Studies}}
\begin{figure}
\setlength{\abovecaptionskip}{0.0cm}
\setlength{\belowcaptionskip}{0.0cm} 
\centering
\subfigure[Yelp]{\includegraphics[width=1.62in]{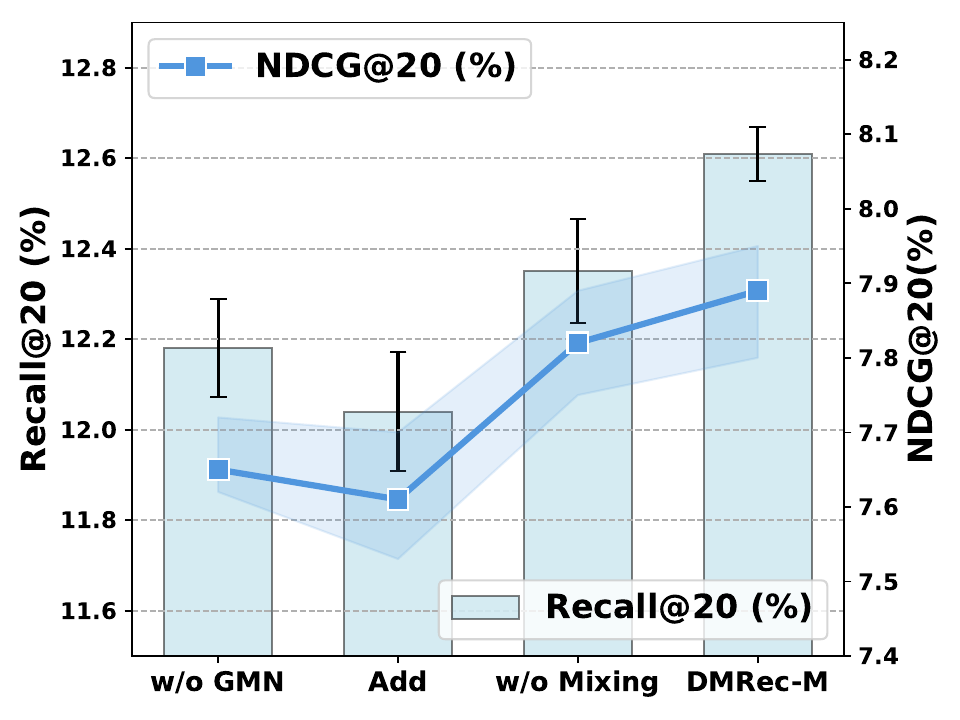}
\label{yelp_ablation_case}}
\hfil
\subfigure[Steam]{\includegraphics[width=1.62in]{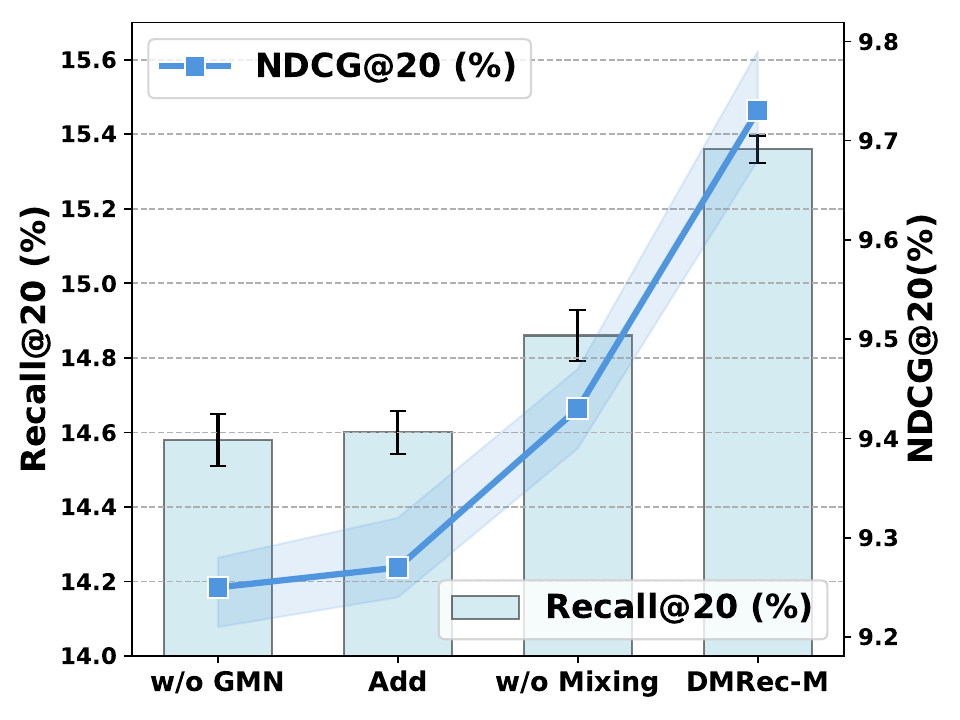}
\label{steam_ablation_case}}

\caption{Ablation studies on (a) Yelp and (b) Steam datasets \textit{w.r.t.} Recall@20 (left $y$-axis) and NDCG@20 (right $y$-axis).}
\label{fig_ablation}
\end{figure}

In this section, we construct several variants to validate the necessity of some of \textsf{DMRec}'s design choices:
\begin{itemize}[leftmargin=*]
\item $\textsf{DMRec}_{\text{w/o PMN}}$: Remove the probabilistic meta-network (Eq. \ref{meta_network}) and directly use MLP to perform the mapping.
\item $\textsf{DMRec}_{\text{Add}}$: Remove the distribution matching strategy and directly add the two distributions for reconstruction.
\item $\textsf{DMRec}_{\text{w/o Mixing}}$: For \textsf{DMRec-M}, remove the mixing design (Eq. \ref{MDDM}) and adopt a joint training process consistent with GODM and CPDM, \textit{i.e.}, $\mathcal L_{\text{w/o mixing}}=\text{D}_{\text{KL}}(q_\phi||p_{\mathbf z}) + \beta \cdot \text{D}_{\text{KL}}(q_\phi||p_{\varphi})$.
\end{itemize}
The experimental results for \textsf{DMRec} and all variants on Yelp and Steam datasets are shown in Fig. \ref{fig_ablation}. After removing the meta-network, the performance of $\textsf{DMRec}_{\text{w/o PMN}}$ drops significantly, indicating that a simple nonlinear mapping is insufficient to establish a connection between the two spaces. The performance of the variant $\textsf{DMRec}_{\text{Add}}$, which simply adds two distributions, is also unsatisfactory. This suggests that merely adding the distributions does not effectively model the cross-space distribution matching. Finally, after removing the mixing mechanism, the performance of $\textsf{DMRec}_{\text{w/o Mixing}}$ generally declines to some extent, indicating the necessity of coordinating the regularization term with the matching term for distribution optimization. Through the mixing mechanism, \textsf{DMRec} can accept distribution information from the language space while maintaining the Gaussian distribution, and simultaneously prevent excessive regularization that could undermine the validity of the posterior.

\subsubsection{\textbf{Hyperparameter Sensitivities}}
\begin{figure}
\setlength{\abovecaptionskip}{0.0cm}
\setlength{\belowcaptionskip}{0.0cm} 
\centering
\subfigure[Yelp]{\includegraphics[width=1.60in]{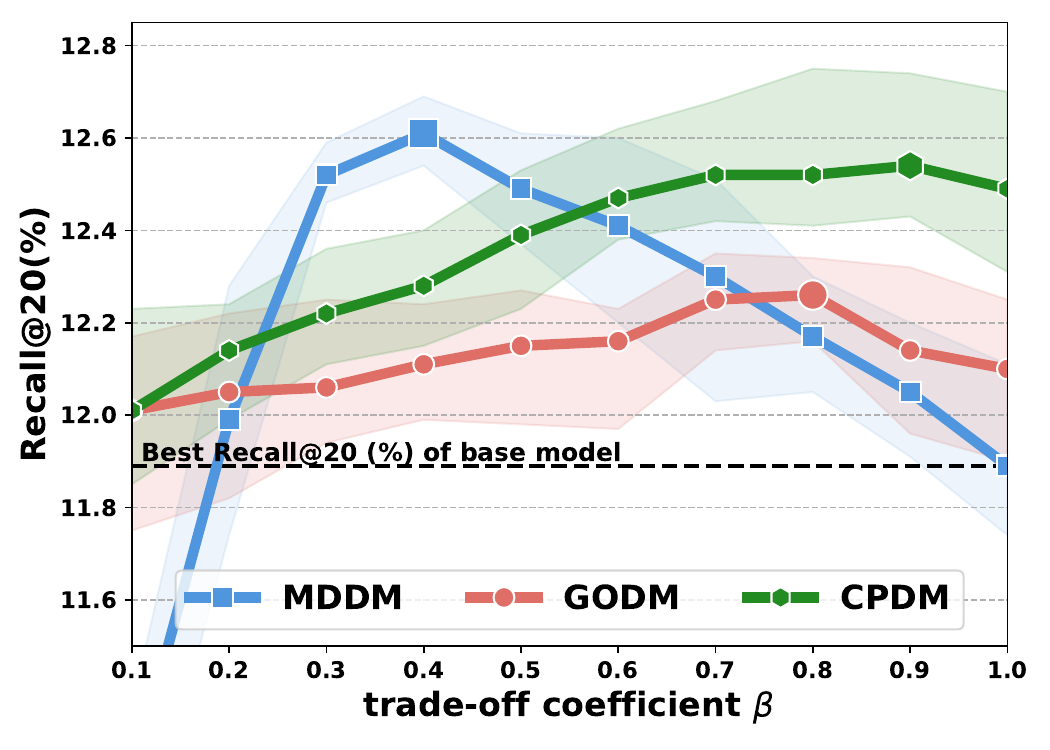}
\label{yelp_param_case}}
\hfil
\subfigure[Steam]{\includegraphics[width=1.60in]{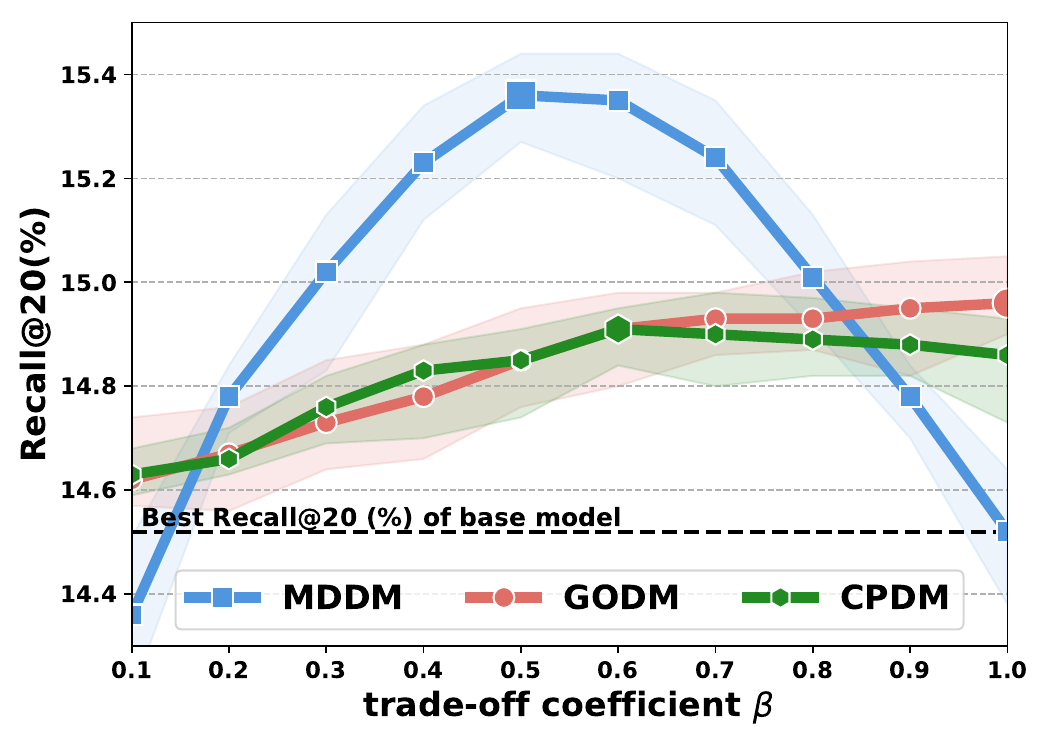}
\label{steam_param_case}}

\caption{Hyperparameter sensitivities for the trade-off coefficient $\beta$ to three matching strategies \textit{w.r.t.} Recall@20 on (a) Yelp and (b) Steam datasets.}
\label{fig_parameter}
\end{figure}

\textsf{DMRec} introduces only one additional hyperparameter $\beta$ to dynamically balance the regularization and matching processes. Fig. \ref{fig_parameter} shows the variation curves of different $\beta$ for three matching strategies based on Mult-VAE. For GODM and CPDM, $\beta$ is applied to the additional matching term, causing the model's performance to increase as $\beta$ starts from 0 and gradually reaches a peak. This indirectly validates the effectiveness of these two matching strategies. For MDDM, which jointly optimizes the regularization and matching terms in Eq. \ref{MDDM}, it exhibits a trend distinct from the other two methods:
\begin{itemize}[leftmargin=*]
\item When $\beta$ is close to 0, it is equivalent to only measuring the matching term. At this point, the performance of \textsf{DMRec} begins to drop sharply and eventually falls below that of the base model. This indicates that relying entirely on the language model significantly disrupts the recommendation process.
\item When $\beta$ is close to 1.0, it is equivalent to only measuring the regularization term. In this case, the performance of \textsf{DMRec} also starts to decline and eventually recovers to the performance of the base model. This suggests that relying solely on the standard Gaussian prior for regularization has a limited effect.
\end{itemize}

\section{Related Work}
\textbf{Generative Model for Recommendation:} The generative model generates a probability over all items and establishes a preference distribution for each user \cite{kingma2013auto, liang2018variational}. It can be mainly categorized into methods based on generative adversarial network (GAN), VAE, and diffusion models \cite{liu2021self, huang2025trustworthiness}. The GAN-based methods \cite{yu2019vaegan, wang2017irgan} train the model through adversarial game-playing, enabling the generator to produce indistinguishable data. In contrast, the VAE-based methods \cite{liang2018variational, zhang2023revisiting} approximate the user distribution by constructing an encoder-decoder structure. Early explorations directly applied the vanilla VAE \cite{kingma2013auto} for recommendation \cite{li2017collaborative}, such as Mult-VAE \cite{liang2018variational}, which introduced a generative model with multinomial likelihood while maintaining the original model structure. Subsequent research has diversified, with examples like BiVAE \cite{truong2021bilateral}, which independently models users and items, and CVGA \cite{zhang2023revisiting}, which uses a graph VAE to explore high-order connectivity. Additionally, techniques such as Gaussian mixture models \cite{xu2020learning}, more complex priors \cite{xu2022alleviating, takahashi2019variational}, and information bottleneck \cite{cao2022cross, wang2025federated} have gradually been incorporated into generative recommendation models. Based on hierarchical VAE \cite{sonderby2016ladder}, diffusion models \cite{ho2020denoising, luo2022understanding} have also been introduced in recommendation scenarios. The most representative work is DiffRec \cite{wang2023diffusion}, which progressively adds Gaussian noise to the original interactions and then removes it during the reverse process. L-DiffRec \cite{wang2023diffusion} improves generalization by shifting the noise addition process to the latent vector space. Although generative methods have made significant progress, the distribution modeling process requires careful attention \cite{higgins2017beta, kingma2016improved}, as it is highly susceptible to disruption, limiting further advancement \cite{wang2022posterior, wang2023diffusion}.

\noindent
\textbf{Language Model for Recommendation:} Due to the powerful text processing capabilities of language model \cite{zhao2023survey}, applying it to recommender systems has received widespread attention \cite{geng2022recommendation, bao2023tallrec, guo2024prompt}. It can be roughly divided into two types: fine-tuned machines \cite{geng2022recommendation} and assistants \cite{ren2024representation}. The first category of research typically integrates the recommender into the fine-tuning of language model \cite{bao2023tallrec, touvron2023llama}. For example, P5 \cite{geng2022recommendation} directly converts interaction data into text prompt, while subsequent works such as TallRec \cite{bao2023tallrec} and LLaRA \cite{liao2024llara} attempt to introduce adapter or LoRA \cite{hulora} for efficient fine-tuning. Overall, the fine-tuning process often requires significant time and computational resources \cite{ren2024representation}, and it is also constrained by the application scenario \cite{kang2018self, liao2024llara}. The second category of methods retains the recommendation model while introducing the language model as an assistant. This process typically involves searching for consistency in the feature space, so transforming text prompts into actionable latent vectors is the primary task of this category of methods \cite{xi2024towards}. Works such as KAR \cite{xi2024towards}, RLMRec \cite{ren2024representation}, and AlphaRec \cite{sheng2024language} point out the rationale and necessity of aligning recommendation model with language model \cite{radford2021learning}. However, these explorations rarely delve into generative models, making them difficult to directly apply to generative recommendation methods.

\section{Conclusion}

In this work, we revisited generative recommendation and proposed a distribution matching-based framework \textsf{DMRec} for generative recommendation. Its core was to model the user preference distribution separately in the collaborative space and the language space. The distribution modeling in collaborative space relied on variational inference, while the distribution modeling in language space was based on the proposed probabilistic meta-network. Subsequently, we proposed three different cross-space distribution matching mechanisms. Empirical experiments on three datasets demonstrated that \textsf{DMRec} could improve the recommendation performance of various types of generative models, while also showing advantages over other language model-enhanced methods.

\begin{acks}
This work is supported by the National Natural Science Foundation of China (No. 62272001), the Australian Research Council under the streams of Future Fellowship (No. FT210100624), the Discovery Early Career Researcher Award (No. DE230101033), the Discovery Project (No. DP240101108 and DP240101814), and the Linkage Projects (No. LP230200892 and LP240200546). 
\end{acks}

\bibliographystyle{ACM-Reference-Format}
\bibliography{sample-base}

% \appendix

\end{document}